\documentclass[prl,twocolumn,superscriptaddress,nofootinbib,amsmath,amssymb]{revtex4-2}

\usepackage{graphicx}
\usepackage{bm}
\usepackage{xcolor}
\usepackage{booktabs}
\usepackage{hyperref}
\usepackage{orcidlink}

\newcommand{\delensalot}[0]{\texttt{delensalot}}
\newcommand{\av}[1]{\left\langle #1 \right\rangle}
\newcommand{\vL}{\boldsymbol{L}}

\begin{document}

\title{De-lensing cosmic birefringence with joint reconstruction}

\author{Sebastian Belkner\,\orcidlink{0000-0003-2337-2926}}
\email{sebastianbelkner@gmail.com}
\affiliation{Universit\'e de Gen\`eve, D\'epartement de Physique Th\'eorique, 24 Quai Ansermet, CH-1211 Gen\`eve 4, Switzerland}
\affiliation{Key Laboratory of Particle Astrophysics, Institute of High Energy Physics (IHEP), Chinese Academy of Sciences, 19B Yuquan Road, Beijing, China}
\affiliation{China Center of Advanced Science and Technology, P.O. Box 8730, Beijing 100190, China}
\author{Julien Carron\,\orcidlink{0000-0002-5751-1392}}
\email{julien.carron@irsol.usi.ch}
\affiliation{Universit\'e de Gen\`eve, D\'epartement de Physique Th\'eorique, 24 Quai Ansermet, CH-1211 Gen\`eve 4, Switzerland}
\affiliation{Department of Physics \& Astronomy, University of Sussex, Brighton BN1 9QH, UK}
\affiliation{Istituto Ricerche Solari Aldo e Cele Dacc\`o (IRSOL), Faculty of Informatics, Universit\`a della Svizzera italiana, 6605 Locarno, Switzerland}

\date{\today}

\begin{abstract}
Reconstruction of anisotropic cosmic birefringence from the CMB is limited by lensing $B$-mode sample variance and partially degenerate with the lensing curl. We study the joint reconstruction of lensing (gradient and curl) and birefringence at the maximum-a-posteriori point, whose spectra are near-optimal statistics.
We predict the joint response and noise analytically, validate them on simulations, and show unbiased inference of its amplitude. Errors will shrink $5$-fold relative to quadratic estimators at SPT-3G+ depths, promising much tighter constraints on parity-violating new physics, owing to the removal of the lensing $B$-polarization sample variance.
\end{abstract}

\maketitle

\emph{Introduction.}---%
Cosmic birefringence, a rotation of the CMB polarization plane as the light travels to us, would indicate parity violation in the electromagnetic sector.
It arises, for instance, from a Chern-Simons coupling of photons to an axion-like field~\cite{Carroll:1989vb,Komatsu:2022nvu}.
Such fields are also candidates for dark matter, dark energy or inflation~\cite{Li:2008tma,Marsh:2015xka}.
In these models the rotation arises because left- and right-handed photons propagate differently.
There is tantalizing evidence from several analyses for a uniform rotation of a fraction of a degree~\cite{Minami:2020odp,Diego-Palazuelos:2022dsq,Diego-Palazuelos:2025dmh,Eskilt:2026imm}.
Confirming this hint requires control of systematic effects degenerate with the signal, in particular the absolute polarization-angle calibration and Galactic foregrounds.
Anisotropic birefringence, a rotation angle $\beta(\hat n)$ varying across the sky, can offer a cleaner window:
it traces fluctuations of the underlying field and induces off-diagonal $EB$ correlations, from which $\beta(\hat n)$ can be reconstructed~\cite{Kamionkowski:2008fp}.
For a scale-invariant spectrum,
\begin{equation}\label{eq:scaleinvariant}
L(L+1)C_L^{\beta\beta} \equiv A_{\rm CB}\,,
\end{equation}
BICEP/Keck data constraint $A_{\rm CB} < 0.044\times 10^{-4}$ (95\% C.L.)~\cite{BICEPKeck:2022kci}, an order of magnitude below power-spectrum-only limits~\cite{Lonappan:2025hwz}, and delensing promises to tighten such constraints further~\cite{IdicherianLonappan:2025trj}.

The most prominent secondary anisotropy of the CMB is gravitational lensing~\cite{Lewis:2006fu}.
Lensing remaps the CMB through a deflection with a gradient mode, now measured at high significance, and a much smaller curl mode sourced by post-Born lens-lens coupling~\cite{Hirata:2003ka,Pratten:2016dsm}.
Deep observations of the CMB can resolve the lensing $B$-modes at the map level, motivating reconstructions beyond quadratic estimators (QE)~\cite{Seljak:1996gy,Hu:1997hv,Hirata:2002jy}. 
Likelihood-based methods approach optimality~\cite{Carron:2017mqf,Millea:2017fyd} and are a key tool to remove the lensing $B$-modes in searches for primordial gravitational waves~\cite{Belkner:2023duz,Aurlien:2022tlp}, for example for the South Pole Observatory program;
the same delensing can benefit birefringence searches.

Three complications arise for deep polarization data.
First, the lensed $B$-mode power acts as a dominant noise for any $EB$ quadratic estimator ($N^{(0)}$);
being a sky signal, it does not decrease with deeper observations, unlike instrumental noise.
Second, lensing also sources a secondary set of connected four-point function contractions, the $N^{(1)}$ bias~\cite{Kesden:2002ku,Hanson:2010rp};
those induced by the very large lensing gradient are sizeable enough to be relevant for birefringence estimation.
Third, the lensing curl and birefringence are both pseudo-scalars and induce partially degenerate $EB$ correlations.
At the QE level the degeneracy can be removed by applying bias hardening techniques~\cite{Namikawa:2012pe}, at the price of extra noise.
Here we show that a joint maximum-a-posteriori (MAP) reconstruction of the lensing gradient $\phi$, the curl $\Omega$, and $\beta$ handles all three at once:
it delenses the polarization, which reduces both $N^{(0)}$ and $N^{(1)}$, and effectively marginalizes the curl-birefringence degeneracy.
The power spectra of the MAP fields are also near-optimal statistics for inferring the field spectra, and exactly optimal in the Gaussian limit, as discussed for example in the appendix of~\cite{Millea:2021had}.
Joint MAP reconstructions have been discussed previously in the context of patchy reionization~\cite{Bianchini:2022wte}, and, closer to this work, in~\cite{Darwish:2025fip}.
Here we derive and validate analytic predictions for the multi-field MAP response and noise, clarify the curl-birefringence degeneracy, and show successful unbiased inference of the birefringence amplitude $A_{\rm CB}$ from the joint MAP reconstructions, demonstrating large gains in signal-to-noise ratio, and in the spherical geometry relevant for large scale signals.

\begin{figure*}
    \centering
    \includegraphics[width=0.9\textwidth]{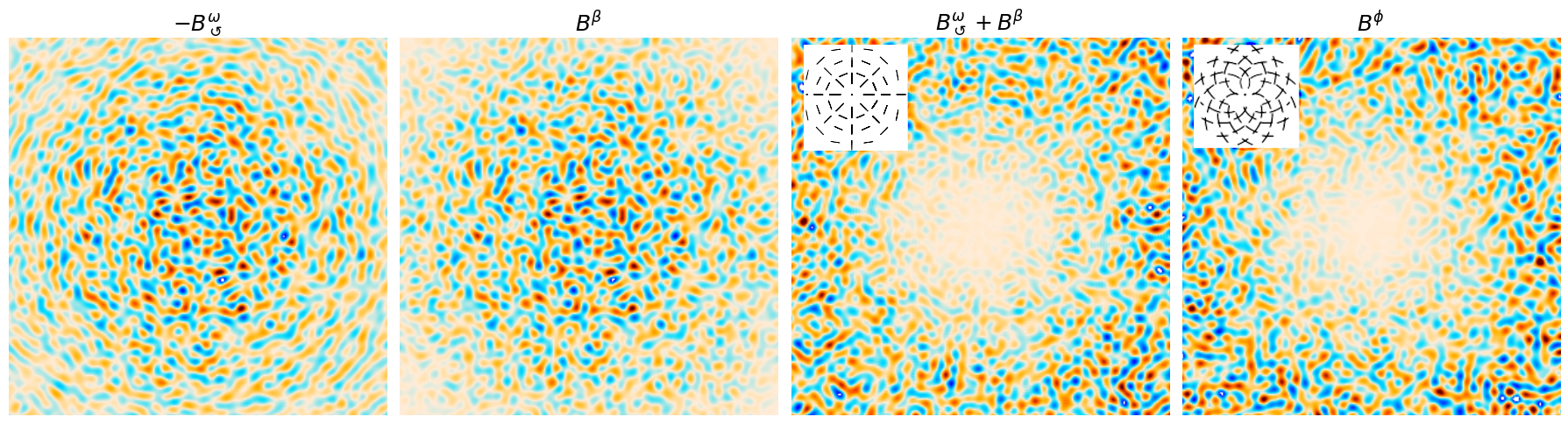}
    \caption{$B$-modes generated from a pure-$E$ field by a single, strongly exaggerated localized source at the patch center, with $\beta = -\omega$.
    \textit{Left to right:} curl-induced $B^\omega$ (display-rotated to align with $B^\beta$); birefringence-induced $B^\beta$; their sum; gradient-lensing $B^\phi$.
    At the center (squeezed limit) the curl rotation and birefringence cancel, leaving only the parity-odd curl shear, which breaks the degeneracy toward the edges.
    Gradient lensing produces no $B$-modes from magnification, only from its shear, rotated by $45^\circ$ relative to the curl shear.}\label{fig:degeneracy}
\end{figure*}

\emph{Curl-birefringence degeneracy.}---%
Lensing remaps the CMB Stokes parameters according to $\tilde X(\hat n) = X(\hat n + \bm\alpha(\hat n))$, with deflection $\bm\alpha = \nabla\phi - \nabla\times\Omega$ (conventions of Ref.~\cite{Carron:2025qxj}).
Locally, the deflection Jacobian decomposes into magnification $\kappa$, shear $\gamma_{ab}$ and rotation $\omega$,
\begin{equation}
\partial_b\alpha_a = -\kappa\,\delta_{ab} - \gamma_{ab} + \omega\,\epsilon_{ab}\,,
\end{equation}
with $\kappa = -\tfrac12\nabla^2\phi$ and a parity-even shear from the gradient mode, and $\omega = -\tfrac12\nabla^2\Omega$, and a parity-odd shear from the curl mode.
Where the Jacobian is approximately constant, the image rotation changes the spin-$\pm2$ polarization $P_{\pm2} = Q\pm iU$ as
\begin{equation}\label{eq:curl_rotation}
\delta P_{\pm2}\big|_{\omega} = \mp 2i\,\omega\, P_{\pm2} + \mathcal O(\omega^2)\,,
\end{equation}
mixing $E$ and $B$ modes locally.
Birefringence rotates the polarization, $P_{\pm2} \to e^{\pm 2i\beta} P_{\pm 2}$, so that
\begin{equation}\label{eq:birefringence_rotation}
\delta P_{\pm2}\big|_{\beta} = \pm 2i\,\beta\, P_{\pm2} + \mathcal O(\beta^2)\,.
\end{equation}
Hence, in this limit the induced change in polarization is the same for the two effects.
The degeneracy is only partial: the curl deflection also carries a parity-odd shear, whose effect depends on its orientation relative to the $E$-mode pattern, while birefringence does not. 
Figure~\ref{fig:degeneracy} illustrates this.
The gradient mode, in contrast, has opposite parity and does not correlate with $\omega$ or $\beta$; since a locally constant magnification does not convert $E$ into $B$, it sources $B$-modes only through its shear.

\emph{Joint MAP reconstruction.}---%
We model the polarization data as $X^{\rm dat} = \mathcal B\,\mathcal D_{\bm\alpha}\,\mathcal F_\beta X + n$, with $X$ the unlensed CMB Stokes polarization, $\mathcal F_\beta$ the birefringence rotation, $\mathcal D_{\bm \alpha}$ the lensing remapping by both $\phi$ and $\Omega$, $\mathcal B$ the beam, and $n$ the noise.
A self-consistent model of birefringence from new physics would be more involved than this separable form, since rotation and lensing build up along the same line of sight~\cite{Namikawa:2021obu,Naokawa:2023upt}, but it is adequate for our purpose of quantifying detection capability.
The reverse ordering, rotation after lensing, is the appropriate one for instrumental systematics such as polarization-angle miscalibrations.
The MAP point $\hat\chi$ maximizes the posterior $p(\chi|X^{\rm dat})$ for the fields $\chi = (\phi, \Omega, \beta)$, with Gaussian priors given by their fiducial spectra.
As in lensing-only previous work~\cite{Belkner:2023duz}, we follow the posterior gradient, except now there are three fields.
For each of these fields, $\chi_a$, $a\in\{\phi,\Omega,\beta\}$, the gradient splits into a quadratic piece, a mean-field and a prior term, $g^{a} = g^{{\rm QD},a} - \left\langle g^{{\rm QD},a}\right\rangle_{\text{fixed $\chi$}} - C_{a}^{-1}\chi_a$.
The quadratic pieces read,
\begin{align}
{}_1g^{\rm QD,\bm\alpha}(\hat n) &= -\sum_{s=\pm2} {}_{-s}\bar X(\hat n)\,\big[\mathcal D_{\bm\alpha}\, \partial_s^+ \mathcal F_\beta\, {}_sX^{\rm WF}\big](\hat n)\,,\\
g^{\rm QD,\beta}(\hat n) &= 4\,{\rm Im}\big[{}_{-2}\bar X(\hat n)\, {}_{+2}(\mathcal D_{\bm\alpha}\mathcal F_\beta X^{\rm WF})(\hat n)\big]\,,
\end{align}
where $\bar X = \mathcal B^\dagger{\rm Cov}^{-1}X^{\rm dat}$ and $X^{\rm WF} = C^{\rm unl}\mathcal F_\beta^\dagger\mathcal D_{\bm\alpha}^\dagger \mathcal B^\dagger {\rm Cov}^{-1}X^{\rm dat}$ are the inverse-variance and Wiener-filtered maps for the current field estimates, and $\partial_s^+$ the spin-raising operator.
The spin-$1$ posterior gradient with respect to the deflection separates into its gradient ($\phi$) and curl ($\Omega$) parts.
These gradients have a simple interpretation~\cite{Legrand:2025vzg}: they are QEs applied to maps delensed and de-rotated by the current estimates, mapped back to the observed coordinates. 
The MAP point is where the QE on the cleaned maps balances the prior.
At the start ($\chi = 0$) they reduce to the standard QEs, which initialize the iterations; we found convergence within 10 iterations.

This picture also predicts the MAP response: with $\hat q_\chi$ the QE-like, prior-free gradient at $\chi$, we expect the linear response to the unresolved residual fields to be $\av{\hat q_\chi} \simeq \bm{\mathcal R}\,(\chi^{\rm true} - \chi)$, where $\bm{\mathcal R}$ is now a $3\times3$ response matrix, which is to be evaluated with partially delensed and de-rotated CMB spectra.
These QEs also have Gaussian noise covariance matrix $\mathbf n^{(0)}$.
Gaussianizing the posterior in these terms gives,
\begin{align}
\hat\chi^{\rm MAP} &\simeq \bm{\mathcal W}^{\rm MAP}\,\hat\chi^{\rm BH}\,,\label{eq:WMAP}\\
\bm{\mathcal W}^{\rm MAP} &= \big(\mathbf C_\chi^{-1} + \mathbf N^{(0),-1}\big)^{-1}\mathbf N^{(0),-1}\,,\nonumber
\end{align}
with $\mathbf N^{(0),-1} = \bm{\mathcal R}^T\mathbf n^{(0),-1}\bm{\mathcal R}$ and $\hat\chi^{\rm BH} = \bm{\mathcal R}^{-1}(\hat q_\chi + \bm{\mathcal R}\chi)$ a set of properly normalized QE estimates.
Joint MAP is thus analogous to jointly Wiener-filtered quadratic estimators.
On modes where the fields are resolved ($\mathbf C_\chi \gg \mathbf N^{(0)}$), it marginalizes the curl-birefringence degeneracy, while elsewhere the prior regularizes the ill-conditioned inversion that can make standard bias hardening noisy.
The $N^{(1)}$ bias follows in the same way, sourced only by the unresolved residual fields. 
Delensing and de-rotating also makes the noise anisotropic~\cite{Legrand:2025vzg}, inducing a small mean-field correction, which we neglect in the reconstruction:
it is about $1\%$ for $\phi$ at $1\,\mu$K-arcmin, and smaller for $\beta$.\footnote{The reason it is so small is that $E$ and $B$ polarization are present in similar amounts in the noise maps -- the anisotropic lensing-like signal induced by distorting the noise is much less relevant than for the E-mode dominated CMB signal.}

\begin{figure}
    \centering
    \includegraphics[width=\columnwidth]{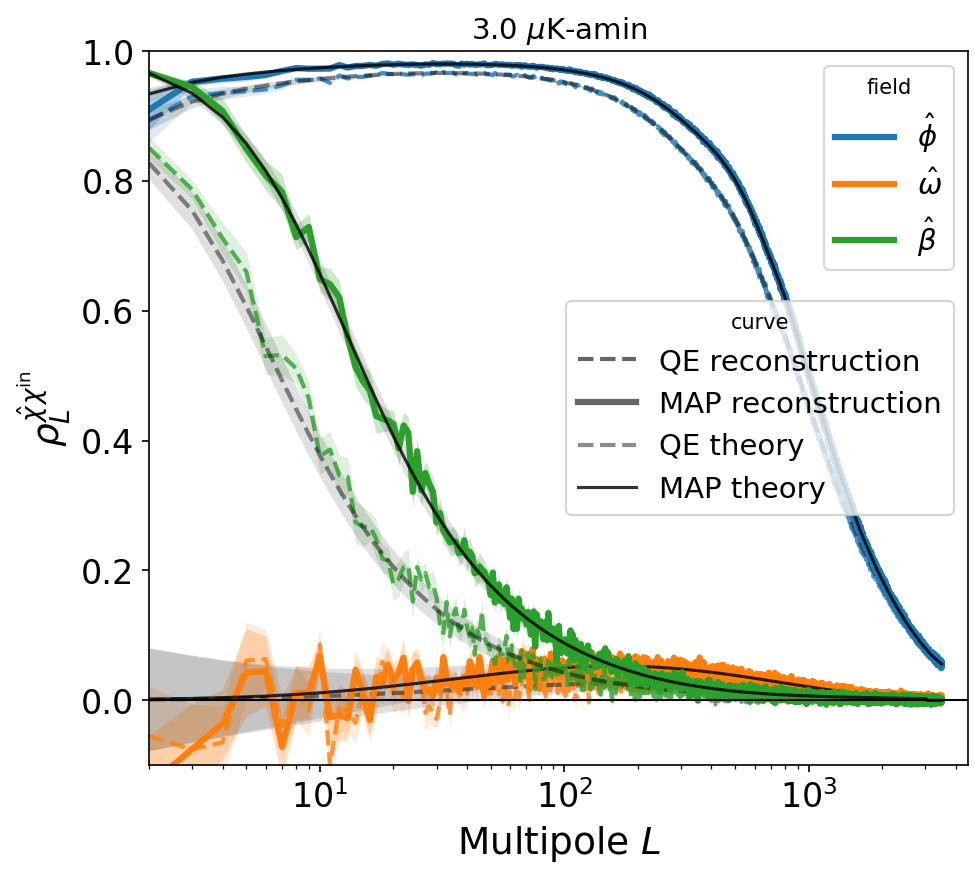}
    \caption{Cross-correlation coefficient between reconstructed and input fields, for QE (dashed) and joint MAP (solid), at $3\,\mu$K-arcmin. Black lines are our predictions including $N^{(0)}$ and $N^{(1)}$; bands show the $1\sigma$ uncertainty on the mean of 32 simulations. MAP improves all fields, most visibly birefringence (green).}
    \label{fig:ccc}
\end{figure}

\emph{Simulations.}---%
We simulate full-sky polarization from Gaussian unlensed $\Lambda$CDM fields with no primordial $B$-modes, rotated by $\beta$ and lensed by $\phi$ and $\Omega$ with \texttt{lenspyx}.
The curl follows the post-Born spectrum~\cite{Robertson:2023xkg}, and $\beta$ is scale-invariant with $A_{\rm CB} = 10^{-7}$.
We add isotropic white noise of $3$, $1.5$ and $1\,\mu$K-arcmin with Gaussian beams of $2$, $1.5$ and $1$ arcmin FWHM, using CMB multipoles $30$ ($E$) or $200$ ($B$) to $4000$, $4500$ and $5000$, and reconstruct the fields for $2\le L \le 3500$, $4000$ and $4500$, respectively.
The $3$ and $1\,\mu$K-arcmin cases are representative of high-resolution SPT-3G~\cite{SPT-3G:2018oot} and SPT-3G+~\cite{SPT:2026olz}.\footnote{These surveys cover only a few percent of the sky. This leads to the same set of complications as for standard QE pipelines. The ratios of QE to MAP errors also are the same.}
We do not model foregrounds; the noise stands for the post-component-separation noise. Each configuration has 32 realizations.
The empirical $3\times3$ MAP response agrees well with the prediction of Eq.~\eqref{eq:WMAP} (Fig.~\ref{fig:response} in the End Matter); cross-responses between fields of opposite parity vanish, as expected.
Our results are stable when removing the curl from simulations and reconstruction, when reversing the order of lensing and rotation, and when changing the beam and $\ell_{\rm max}$.
Quadrupling the simulated $A_{\rm CB}$ while keeping the fiducial prior leaves the MAP Wiener filter unchanged within $3\%$.

\emph{Results.}---%
Figure~\ref{fig:ccc} shows that MAP reconstructs every field better than the QE, in good agreement with our predictions.
The gain comes mostly from the reduced lensing $B$-mode power in the $N^{(0)}$ noise of $\omega$ and $\beta$, and from the strong suppression of the $\phi$-induced $N^{(1)}$.
For the birefringence QE, $N^{(1)}$ is about $20\%$ of the noise at $3\,\mu$K-arcmin and almost $50\%$ at $1\,\mu$K-arcmin; for MAP it drops to about $5\%$ at all noise levels.

\begin{figure}
    \centering
    \includegraphics[width=\columnwidth]{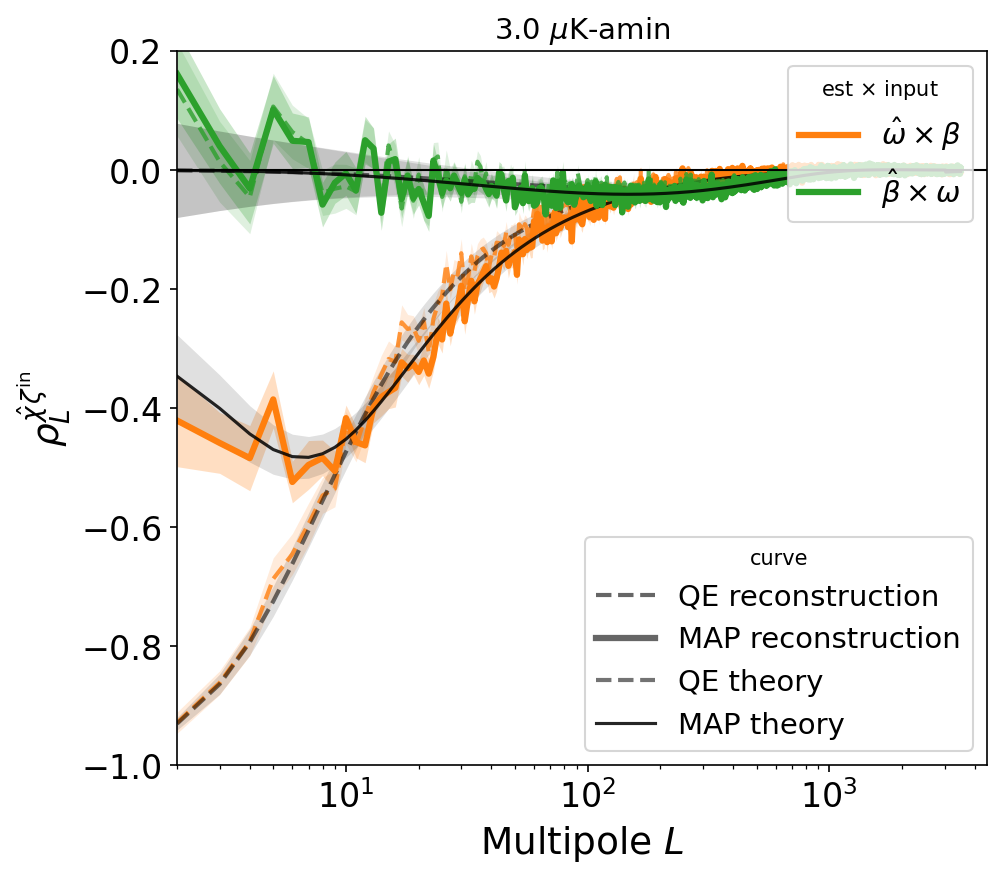}
    \caption{Cross-correlation coefficients between reconstructed and input $\omega$ and $\beta$ fields at $3\,\mu$K-arcmin, for QE (dashed) and joint MAP (solid), with the simulation scatter as bands and our predictions in black. The reconstructed curl correlates with the input birefringence (orange) through the degeneracy; joint MAP strongly reduces this where birefringence is resolved ($L \lesssim 10$). Full bias hardening would set these curves to zero, at the price of a larger variance.}
    \label{fig:ccin}
\end{figure}

Figure~\ref{fig:ccin} shows the curl-birefringence degeneracy.
The curl is noise dominated at all scales, so the reconstructed $\beta$ hardly correlates with the input curl.
The opposite does not hold: our scale-invariant birefringence is large on large scales, and the reconstructed curl correlates strongly with the input $\beta$ for both estimators.
On scales where every birefringence mode is resolved ($L\lesssim 10$--$20$), joint MAP removes much of this correlation, as expected from the marginalization in Eq.~\eqref{eq:WMAP}.

\begin{figure}
    \centering
    \includegraphics[width=0.99\columnwidth]{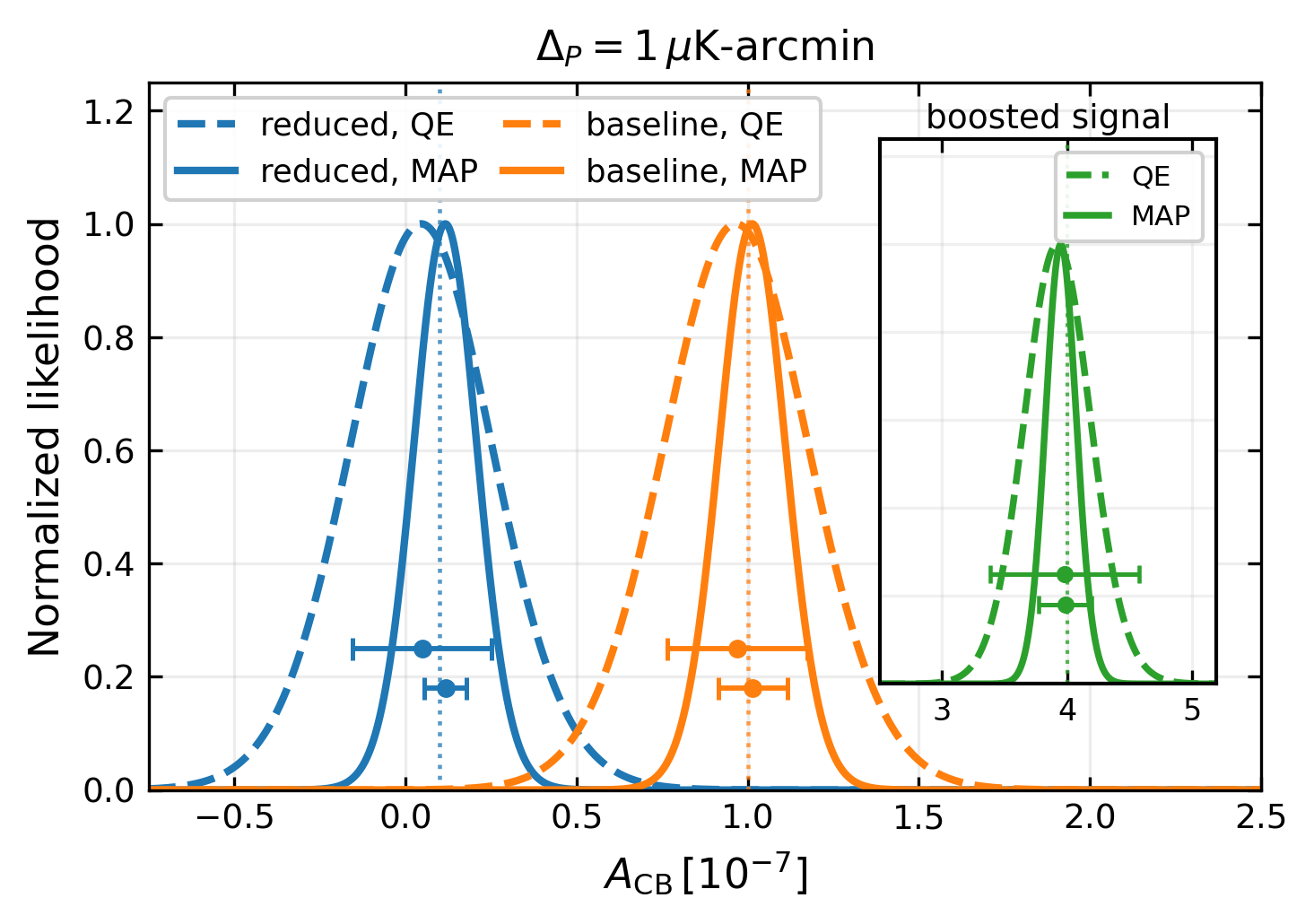}
    \caption{Example posteriors for the birefringence amplitude at $1\,\mu$K-arcmin, for QE (dashed) and joint MAP (solid), with input amplitudes $A_{\rm CB} = 10^{-8}$ (blue), $10^{-7}$ (orange, baseline) and $4\times10^{-7}$ (green, inset), marked by dotted lines, using birefringence multipoles down to $L\sim 100$, reconstructed from CMB modes down to $\ell\sim 4500$.
    Thick lines show the log-average of the per-simulation posteriors, peak-normalized.
    Points with error bars show the mean and scatter of the posterior means across simulations.
    All likelihood ingredients, including the covariance, are taken from the baseline simulations and kept fixed across input amplitudes. The posterior widths are therefore calibrated only at the baseline amplitude, while the error bars give the actual uncertainty at each amplitude.
    Neither estimator shows a significant bias. The MAP gain over QE is largest where the cosmic variance contribution is smallest, at low $A_{\rm CB}$.}
\label{fig:likelihood}
\end{figure}

To constrain $A_{\rm CB}$, we use the birefringence auto-spectrum of the joint reconstructions.
Since the MAP response is a matrix in field space, we compare QE and MAP in their respective Wiener-filtered spaces, where the birefringence estimate is the linear combination of the reconstructed fields with minimum error variance.
For QE we filter the measured spectra as $\hat S_L^{\rm QE} = [\bm{\mathcal W}^{\rm QE}_L\hat{\mathbf C}_L\bm{\mathcal W}^{{\rm QE},T}_L]_{\beta\beta}$, while the MAP reconstruction is already filtered, $\hat S_L^{\rm MAP} = \hat C_L^{\beta\beta}$.
We subtract the noise biases, $N^{(0)}$ and the lensing-induced $N^{(1)}$, from the measured spectra and compare the result to $[\bm{\mathcal W}^X_L(\mathbf C_L + \mathbf N^{(1),\beta}_L)\bm{\mathcal W}^{X,T}_L]_{\beta\beta}$ for $X \in \{{\rm QE}, {\rm MAP}\}$, with $\mathbf C_L$ the fiducial field spectra with birefringence amplitude $A_{\rm CB}$, and $\mathbf N^{(1),\beta}_L$ the $N^{(1)}$ sourced by birefringence, which scales with $A_{\rm CB}$.
The QE filter $\bm{\mathcal W}^{\rm QE}_L$ is built from the standard analytic QE noise.
For MAP, we effectively calibrate percent-level effects on the filter and noise term with simulations: $\bm{\mathcal W}^{\rm MAP}_L$ from the cross-spectra of the baseline reconstructions with their input fields, and the noise term from simulations without birefringence, except for the birefringence-sourced $N^{(1)}$ (see End Matter).
We bin these spectra, estimate their covariance from the baseline simulations, and adopt the Sellentin-Heavens likelihood~\cite{Sellentin:2015waz}.
Figure~\ref{fig:likelihood} shows the resulting posteriors at $1\,\mu$K-arcmin. Both estimators show no significant bias for all configurations, and the results are stable against the binning.
They remain similarly unbiased when the data are far from our fiducial point, for example for an input signal four times larger, $A_{\rm CB} = 4\times10^{-7}$, with the MAP prior and the likelihood ingredients unchanged.

\begin{table}
\caption{Improvement $\sigma_{\rm QE}/\sigma_{\rm MAP}$ of the constraint on $A_{\rm CB}$, full sky: empirical (scatter of the likelihood peaks over 32 simulations, uncertainty $\sim\!20\%$), predicted (binned Fisher), and predicted without birefringence signal ($A_{\rm CB} = 0$), the relevant case for upper limits.}
\label{tab:improvement}
\begin{ruledtabular}
\begin{tabular}{cccc}
Noise [$\mu$K-arcmin] & empirical & predicted & predicted, $A_{\rm CB}=0$ \\
\hline
$3.0$ & $1.64$ & $1.32$ & $1.71$ \\
$1.5$ & $2.05$ & $1.76$ & $3.21$ \\
$1.0$ & $2.38$ & $2.17$ & $4.97$ \\
\end{tabular}
\end{ruledtabular}
\end{table}

Table~\ref{tab:improvement} summarizes the gain.
With our fiducial signal, MAP tightens the constraint on $A_{\rm CB}$ by up to a factor $2.4$, in agreement with the prediction within the simulation scatter.
The gain grows with decreasing noise, as more of the lensing $B$-modes are removed.
A significant part of this error comes from $\beta$ cosmic variance.
Without $A_{\rm CB}$ signal, the predicted improvement reaches a factor of almost $5$ at $1\,\mu$K-arcmin.

\emph{Outlook.}---%
Joint MAP reconstruction of lensing and anisotropic birefringence tightens birefringence constraints by up to a factor $\sim\!5$ for deep forthcoming observations, mostly by delensing, through the reduction of both $N^{(0)}$ and $N^{(1)}$.
The curl-birefringence degeneracy is marginalized, with little impact for a scale-invariant birefringence spectrum.
The impact of marginalization can be larger for other spectral shapes; our joint reconstruction and analytic prediction is designed to handle these equally well.
Our analytic predictions for the multi-field MAP response and noise, validated on map-level reconstructions, can be used for inference without significant bias.
A key element making this possible is the fact that the response matrix of the MAP point is essentially independent of the data.
We have not modelled Galactic foregrounds: spatially varying dust could produce off-diagonal $EB$ correlations, but dust power decays sharply with multipole, and there is no particular reason to expect that they should significantly impact our results, which rely on small CMB scales.
The same framework extends to other secondaries with a practical enough forward model, such as patchy reionization, or to instrumental systematics such as polarization-angle miscalibration.

\begin{acknowledgments}
The authors thank Omar Darwish for discussions early in the project that strongly shaped this work.
SB thanks Hongbo Cai for fruitful discussions.
SB further acknowledges support from Claude and ChatGPT in plotting and debugging code.
JC and SB acknowledge support from a SNSF Eccellenza Professorial Fellowship (No. 186879).
This work was supported by the International Postdoctoral Fellowship of the Institute of High Energy Physics, Chinese Academy of Sciences (No.~E65153U110).
The code is available in \delensalot\footnote{\url{https://github.com/NextGenCMB/delensalot}}.
\end{acknowledgments}

\bibliography{references}

\onecolumngrid
\section*{End Matter}
\emph{Likelihood details}---%
Our likelihood for $A_{\rm CB}$ relies on analytic predictions for the MAP response matrix (`Wiener filters'), and for the $N^{(0)}$ and $N^{(1)}$ biases. We give further details here.

Exact analytic agreement of the responses and other fiducial ingredients is in fact not required for the likelihood.
What matters is that they do not depend on the birefringence signal in the data, but only on fiducial ingredients, such that they can be calibrated on simulations generated from these fiducial elements.
In practice, for MAP we take both from simulations:
the Wiener filter is measured on the baseline simulations, and the noise term, $N^{(0)}$ and the lensing-induced $N^{(1)}$, is the mean reconstructed spectrum of simulations with $A_{\rm CB}=0$, without theory input.
Only the $N^{(1)}$ sourced by birefringence, which scales with $A_{\rm CB}$, is taken from theory.
All of these, as well as the covariance, are kept fixed for all our $A_{\rm CB}$ input values.
Using analytic predictions exclusively would instead bias our MAP posteriors high by about $0.6\sigma$ for $A_{\rm CB} = 10^{-7}$ and $4\times10^{-7}$, compared to $0.2\sigma$ with the measured ingredients.
For the QE, the analytic predictions are sufficient, and we use them throughout.

More generally, analytic modelling of the likelihood ingredients is simplified because for signals as small as considered here, the rotation itself barely changes the CMB spectra. For our fiducial $A_{\rm CB}=10^{-7}$, the $B$-mode power produced by rotation is about $10^{-4}$ of the lensing $B$-mode power.
The fractional change in the $E$-mode power is also negligible, so the spectra entering the weights, the responses and the biases can be taken without rotation.
Further, even though our likelihood approach uses all information, only the $EB$ part of the rotation estimators (`likelihood gradients') really matters at low $L$: in the absence of primordial $B$-modes, a large-scale rotation mixes $E$ and $B$ but leaves the $EE$ power unchanged at first order, by Eq.~\eqref{eq:birefringence_rotation}.
The $EE$ response therefore vanishes as $L\to0$, and the $EE$ part of the estimator is negligible there. 

\begin{figure}[h]
    \centering
    \includegraphics[width=0.7\textwidth]{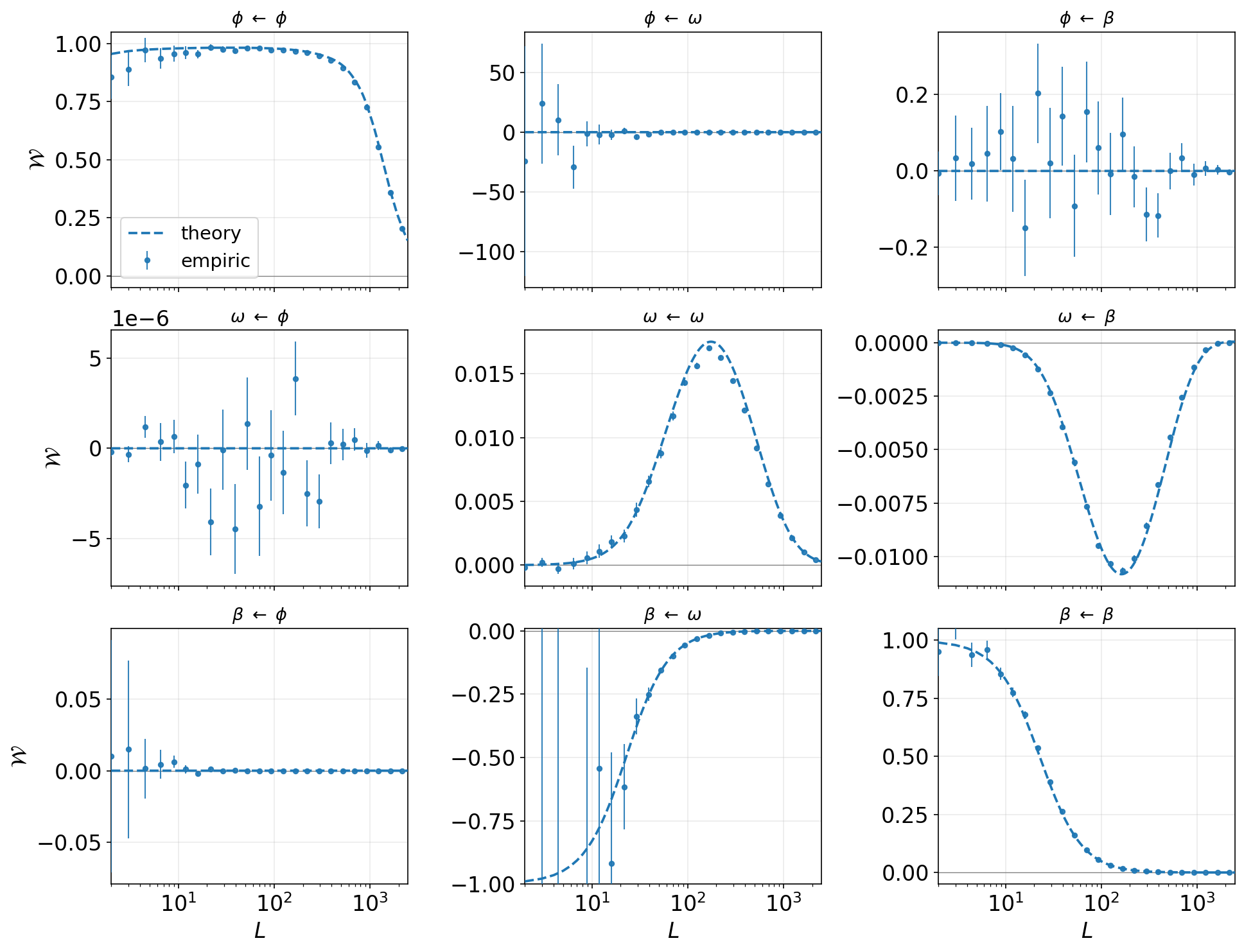}
    \caption{Joint MAP response matrix $\mathcal W^{ab}_{L,\rm emp}$ (Eq.~\eqref{eq:Wemp}) for the $1\,\mu$K-arcmin configuration: empirical estimates (blue points, $1\sigma$ errors from 32 simulations) and the linearized prediction including the noise-induced mean-field correction (lines). Rows and columns correspond to $\phi$, $\omega$ and $\beta$. Diagonal panels show the Wiener-filter response of each field, off-diagonal panels the cross-responses. Cross-responses between parity-even ($\phi$) and parity-odd ($\omega$, $\beta$) fields vanish.}
    \label{fig:response}
\end{figure}

We compare the linearized prediction for the joint MAP response, Eq.~\eqref{eq:WMAP}, with an empirical measurement. For uncorrelated input fields, the response of the reconstructed field $\hat\chi_a$ to the input field $\chi_b^{\rm in}$ is estimated as
\begin{equation}\label{eq:Wemp}
\mathcal W^{ab}_{L,\rm emp} \equiv \frac{\hat C_L^{\hat\chi_a\chi_b^{\rm in}}}{C_L^{\chi_b^{\rm in}\chi_b^{\rm in}}}\,.
\end{equation}
Figure~\ref{fig:response} shows the full $3\times3$ matrix for the $1\,\mu$K-arcmin configuration. The agreement with the predictions is very good. Responses between fields of opposite parity, $\phi$ to $(\omega, \beta)$ and vice versa, vanish by parity selection rules, and are consistent with zero in our simulations.
The $\beta$ response to the input curl has the opposite sign to the birefringence auto-response, as expected from the degeneracy for $\beta = -\omega$.
It is noisy on the largest scales, owing to the limited number of simulations and the low signal-to-noise of the curl there.

The theoretical responses entering the response matrix are built from the response functions $f^{\chi XY}_{\ell_1\ell_2L}$ of a pair of CMB fields to an anisotropy source $\chi \in \{\phi, \Omega, \beta\}$, and from the estimator weights $g^{xXY}_{\ell_1\ell_2L}$, which include the inverse-variance filtering~\cite{Okamoto:2003zw}. The response of estimator $x$ to the source $\chi$ is
\begin{equation}\label{eq:Rth}
\mathcal R^{x\chi}_L = \frac{1}{2(2L+1)}\sum_{\ell_1\ell_2} g^{xXY}_{\ell_1\ell_2L}\,f^{\chi XY}_{\ell_1\ell_2L}\,,
\end{equation}
with an implied sum over the field pairs $XY$. In the absence of primordial $B$-modes, the $EB$ response functions are
\begin{equation}\label{eq:fcurved}
f^{\phi EB}_{\ell_1\ell_2L} = \frac{i}{2}\left[L(L+1) + \ell_1(\ell_1+1) - \ell_2(\ell_2+1)\right]C^{EE}_{\ell_1}\,\mathcal N_{\ell_1\ell_2L}\;{}_2F^{-}_{L\ell_1\ell_2}\,,\qquad
f^{\beta EB}_{\ell_1\ell_2L} = -2\,C^{EE}_{\ell_1}\,\mathcal N_{\ell_1\ell_2L}\;{}_2F^{+}_{L\ell_1\ell_2}\,,
\end{equation}
where $\mathcal N_{\ell_1\ell_2L} = \sqrt{(2\ell_1+1)(2\ell_2+1)(2L+1)/4\pi}$ and ${}_2F^{\pm}_{L\ell_1\ell_2} = \tfrac12\big[\big(\begin{smallmatrix}L&\ell_1&\ell_2\\0&-2&2\end{smallmatrix}\big) \pm \big(\begin{smallmatrix}L&\ell_1&\ell_2\\0&2&-2\end{smallmatrix}\big)\big]$.

We compute the $N^{(1)}$ biases in the flat-sky approximation, following Ref.~\cite{Planck:2018lbu}. With $\bm\ell_1+\bm\ell_2 = \vL$, the flat-sky response functions are defined by $\av{X_{\bm\ell_1}Y_{\bm\ell_2}} \ni f^{\chi XY}_{\bm\ell_1\bm\ell_2}\,\chi_{\vL}$ at fixed CMB spectra. For the $EB$ pair, with $\varphi_{12}$ the angle between $\bm\ell_1$ and $\bm\ell_2$, the counterparts of Eq.~\eqref{eq:fcurved} are
\begin{equation}\label{eq:fflat}
f^{\phi EB}_{\bm\ell_1\bm\ell_2} = -(\vL\cdot\bm\ell_1)\,C^{EE}_{\ell_1}\sin2\varphi_{12}\,\quad
f^{\beta EB}_{\bm\ell_1\bm\ell_2} = 2\,C^{EE}_{\ell_1}\cos2\varphi_{12}.
\end{equation}
 For estimators $\hat x_{\vL} = (\mathcal R^{xx}_L)^{-1}\int_{\bm\ell_1} g^{xXY}_{\bm\ell_1\bm\ell_2}\,X_{\bm\ell_1}Y_{\bm\ell_2}$, with weights $g^{xXY}_{\bm\ell_1\bm\ell_2}$ that include the inverse-variance filtering and $\int_{\bm\ell} \equiv \int d^2\ell/(2\pi)^2$, the bias is
\begin{equation}\label{eq:n1flat}
N^{(1),xy}_{L} = \frac{1}{\mathcal R^{xx}_L\mathcal R^{yy}_L}\int_{\bm\ell_1}\int_{\bm\ell_1'} g^{xXY}_{\bm\ell_1\bm\ell_2}\,g^{yZW}_{\bm\ell_1'\bm\ell_2'}
\sum_\chi\Big[C^{\chi\chi}_{|\bm\ell_1+\bm\ell_1'|}\,f^{\chi XZ}_{\bm\ell_1\bm\ell_1'}\,f^{\chi YW}_{\bm\ell_2\bm\ell_2'} + C^{\chi\chi}_{|\bm\ell_1+\bm\ell_2'|}\,f^{\chi XW}_{\bm\ell_1\bm\ell_2'}\,f^{\chi YZ}_{\bm\ell_2\bm\ell_1'}\Big],
\end{equation}
with $\bm\ell_2 = \vL-\bm\ell_1$ and $\bm\ell_2' = -\vL-\bm\ell_1'$.

For birefringence or other spin-0 estimators, an exact curved-sky check is available. The scale-invariant signal peaking at low $L$ places our likelihood well in the squeezed limit, $L \ll \ell$, where the bias is close to its value at the monopole $L=0$. The $N^{(1)}$ bias of the normalized estimator induced by an anisotropy source $\chi \in \{\phi, \Omega, \beta\}$ then collapses to (see the appendix of Ref.~\cite{Carron:2025qxj})
\begin{equation}\label{eq:n1L0}
N^{(1),\beta\beta}_{L=0} = \frac{1}{4\big(\mathcal R^{\beta\beta}_{L=0}\big)^2} \sum_{\ell_1\ell_2 L'} \left(\frac{(-1)^{\ell_1} g^{\beta XY}_{\ell_1\ell_1 0}}{\sqrt{2\ell_1+1}}\right)\left(\frac{(-1)^{\ell_2} g^{\beta ZW}_{\ell_2\ell_2 0}}{\sqrt{2\ell_2+1}}\right) C^{\chi\chi}_{L'}
\left[(-1)^{P} f^{\chi XW}_{\ell_1\ell_2L'}f^{\chi YZ}_{\ell_1\ell_2L'} + (-1)^{P'} f^{\chi XZ}_{\ell_1\ell_2L'}f^{\chi YW}_{\ell_1\ell_2L'}\right].
\end{equation}
Here a sum over the field pairs $XY$ and $ZW$ entering the estimator is implied, and $\mathcal R^{\beta\beta}_{L=0}$ is the birefringence response of Eq.~\eqref{eq:Rth} at $L=0$. The signs $(-1)^P$ and $(-1)^{P'}$ are the parities of the three-field combinations $(X, W, \chi)$ and $(X, Z, \chi)$.

At $1\,\mu$K-arcmin, we find a $\phi$-induced $N^{(1)}_{L=0}/N^{(0)}_{L=0}  \sim 0.5$ for the QE but dropping to a couple of percent for MAP, using partially delensed spectra and the residual lensing power. In our configurations, neglecting $N^{(1)}$ in the likelihood model can significantly bias the QE inference of $A_{\rm CB}$, whereas for MAP the shift remains within $1\sigma$.
\end{document}